\documentclass[prb,twocolumn,showpacs,superscriptaddress, amsmath,amssymb]{revtex4}

\usepackage{graphicx}
\usepackage{dcolumn}
\usepackage{bm}
\usepackage[dvips]{color}
\usepackage{color}

\newcommand{\vect}[1]{\mathbf{#1}}

\newcommand{\rfm}{$\rm RbFe(MoO_4)_2$}

\begin{document}
\title{Anisotropy stabilized magnetic phases of the triangular antiferromagnet RbFe(MoO$_4$)$_2$}

\author{Yu.~A.~Sakhratov}
\affiliation{National High Magnetic Field Laboratory, Tallahassee, Florida 32310, USA}
\affiliation{Kazan State Power Engineering University, 420066 Kazan, Russia}
\author{L.~E.~Svistov}
\email{svistov@kapitza.ras.ru}
\affiliation{P.L. Kapitza Institute for Physical Problems, RAS, Moscow 119334, Russia}
\author{A.~P.~Reyes}
\affiliation{National High Magnetic Field Laboratory, Tallahassee, Florida 32310, USA}
\date{\today}
\begin{abstract}
The magnetic  $H-T$ phase diagram of a quasi-two-dimensional easy plane antiferromagnet \rfm\ ($S=5/2$) with an equilateral triangular lattice structure is studied with $^{87}$Rb NMR technique for field  directed along hard axis $C_3$. The studies confirm the two step transition from the low field umbrella-like incommensurate magnetic phase to the paramagnetic state observed recently (Mitamura et al. 2016). The transitions were identified as a lambda anomaly in the spin-lattice relaxation rate and a step increase of magnetic susceptibility at intermediate transition. The $^{87}$Rb NMR study precludes the possibility of either V or fan spin textures in the X phase. The additional transition is presumably associated with loss of inter plane magnetic order before the transition to paramagnetic state of individual triangular planes.
\end{abstract}
\maketitle
\section{Introduction} \label{Intro}

\rfm\ is an example of a quasi-two-dimensional (2D) antiferromagnet ($S=5/2$) with a regular triangular lattice structure.
In this compound the antiferromagnetic inter-planar exchange interaction $J'$ is much weaker than the dominant in-plane interaction $J$
($J'/J\approx 0.01$). This defines the quasi-two dimensional character of the magnetic state in \rfm.
For a magnetic field applied within the triangular plane, a sufficiently strong single-ion easy-plane anisotropy leads to a strong resemblance between the main features of the $H-T$ phase diagram for \rfm~\cite{Inami_1996, Svistov_2003, Kenzelmann_2004, Svistov_2006, Kenzelmann_2007, Smirnov_2007, White_2013, Mitamura_2014, Zelenskiy_2021} and the model phase diagram of the $XY$ 2D triangular lattice antiferromagnet (TLAF) studied in detail in Refs.~[\onlinecite{Korshunov_1986, Lee_1986, Chubukov_1991, Gekht_1997}].
The magnetic states of TLAF model system are the infinitely degenerate ground states found in the quasi classical limit of large spins on the lattice nodes. As a result the magnetic structures in the field-temperature (H-T) phase diagram is expected to be defined either by week magnetic interactions or quantum and thermal fluctuations.
The magnetic structures of the XY 2D TLAF with increasing field applied along the triangular plane are sketched in Fig. 1a. The magnetic structures depicted in the figure as "Y","UUD" and "V"are realized in the wide field range $0.14H_{sat}\lesssim H \lesssim 0.7H_{sat}$. The spin arrangement in each triangular plane of these   phases coincide with the prediction of the $XY$ 2D TLAF model.
The dominant structure that arises in  model is ultimately determined by the quantum and thermal fluctuations. This phenomena heightened the scientific interest in studying  \rfm\ as a magnet, in which the ordering scenario called "order by disorder" is realized. A review of experimental data concerning magnetic structures of \rfm\ obtained with different techniques at field applied along the triangular plane and the description of high field magnetic phase is given in Ref.~[\onlinecite{LAST}].

In case of sufficiently strong "easy plane" anisotropy, the umbrella-like magnetic structure ("U"-structure) is expected in 2D TLAF model up to saturation field when applied perpendicular to triangular planes of \rfm. The sketch of this structure with increasing field are shown in the Fig.1b. The projections of magnetic moments of sublattices on triangular plane are 120 degrees apart. The projections of sublattices on field axis equal and increases from zero at $H=0$ to the polarized state at $H_{sat}$. In the frame of this model the rotation of the 2D structure around $C_3$ does not change magnetic energy. The magnetic structure of \rfm\ for $H=0$  was studied with elastic neutron scattering. The wave vector of magnetic structure was found equal to (1/3,~1/3,~0.458)~\cite{Kenzelmann_2007}. The sublattices of the neighbor planes rotate at an angle of 165 degrees around $C_3$. This incommensurate arrangement can be explained by antiferromagnetic exchange interactions of magnetic sublattices of neighbor planes
considered in Ref.~[\onlinecite{Zelenskiy_2021}]. Magnetometry study in pulse fields $H\parallel C_3$ of \rfm~\cite{Smirnov_2007} did not detect any peculiarity in the fields up to saturation. This leads to the assumption that the "U" structure only takes place during magnetization. The later studies of elastic and magnetoelectric properties of \rfm~\cite{Mitamura_2016, Zelenskiy_2021} show that the phase diagram for this field direction is more complex. In a wide field range below saturation a new magnetic phase was found. For simplicity, this phase will be denoted as X-phase. The X-phase, in contrast to U-phase demonstrates no electric polarization. The transition from U-phase to X-phase is accompanied by anomalies on sound velocity. It was suggested \cite{Mitamura_2016, Zelenskiy_2021}, that in X-phase the planar V-structure is realized similar to V-structure for field applied in triangular plane (Fig. 1a).

In this paper we study the magnetic structure  of \rfm\ in fields up to saturation, directed perpendicular to triangular planes with use of $^{87}$Rb NMR technique. The NMR spectra were compared with results of modeling in frame of U and V structures.

 \begin{figure}[tb]
\includegraphics[width=0.85\columnwidth]{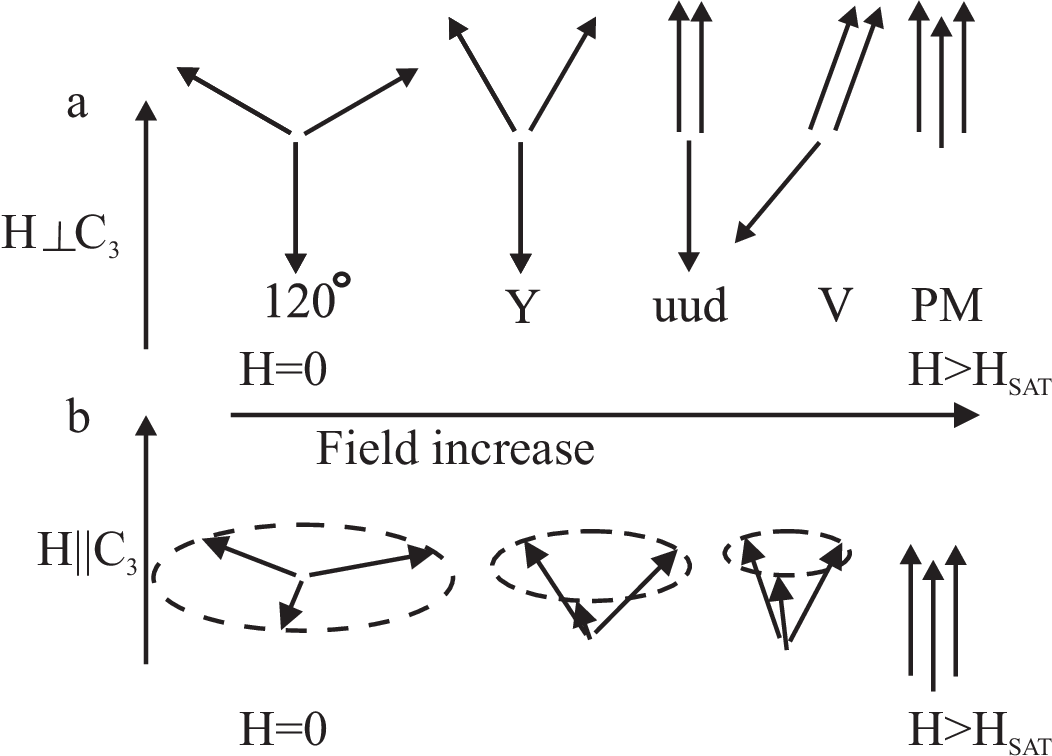}
\caption{
(a) Magnetic structures expected for the XY TLAF model which includes spin fluctuations for the increasing magnetic field  $H\perp C_3$
(b) "Umbrella like" magnetic structures expected for the TLAF model with easy plane anisotropy in quasi classical model for the increasing magnetic field $H\parallel C_3$}
\label{fig1_structures}
\end{figure}

Prior to describing the experimental procedures and presenting our results, in the next section we review the magnetic properties of \rfm.

\section{Magnetic interactions and structure of \rfm}

\begin{figure}[tb]

\includegraphics[width=0.85\columnwidth]{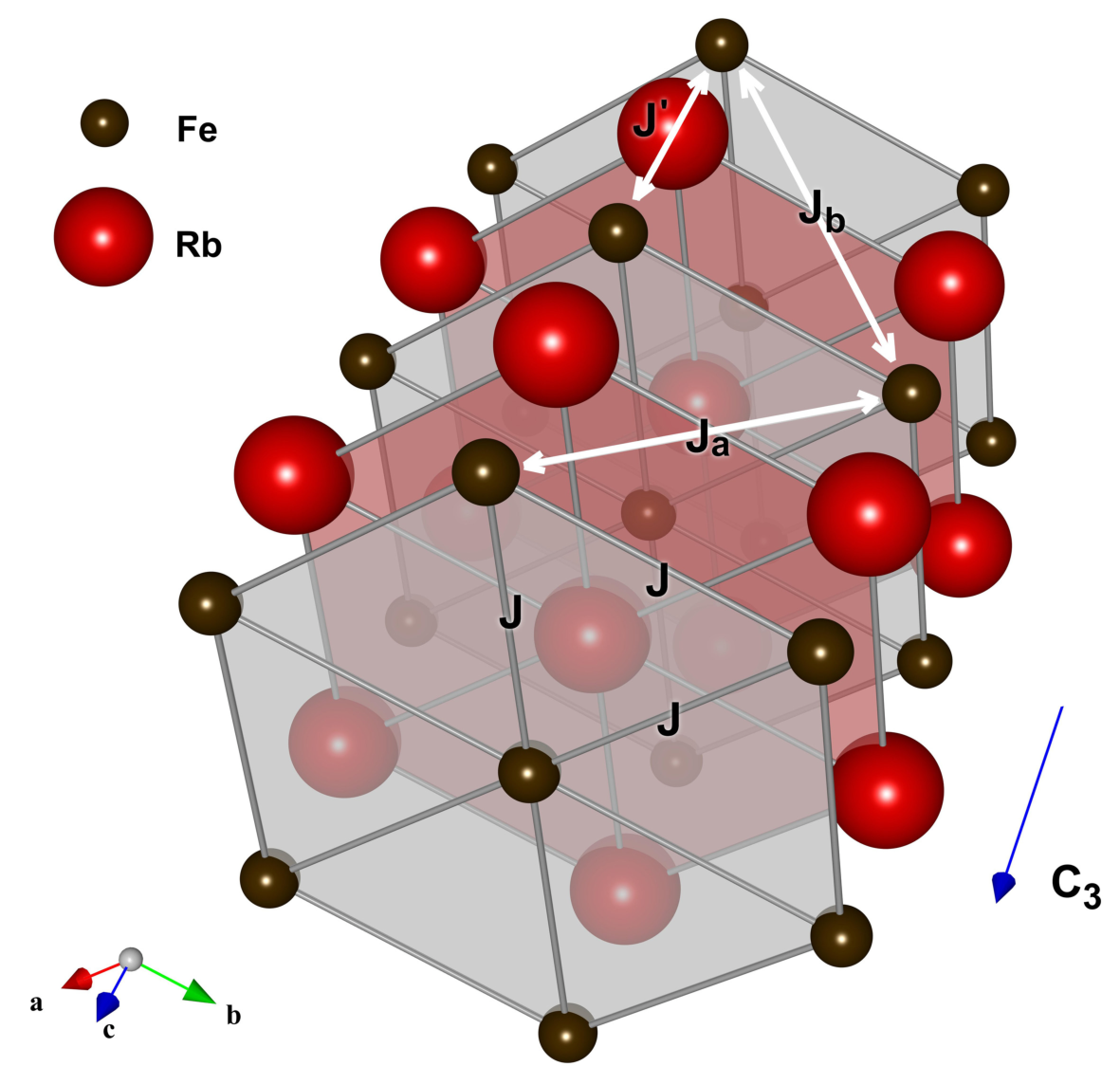}
\vspace{0 mm}
\caption{Crystal structure of \rfm.
The smaller black spheres mark the positions of the Fe$^{3+}$ magnetic ions, the larger red spheres are at the positions of the Rb$^+$ ions, while the (MoO$_4$)$^{2-}$ complexes are not shown.
The crystal symmetry dictates the following main exchange bonds within one cell: in-plane bonds $J$, inter-plane bonds $J'$, $J_a$, $J_b$.}
\label{fig2_crystal}
\end{figure}

The crystal structure of \rfm\ consists of alternating layers of Fe$^{3+}$, (MoO$_4$)$^{2-}$, and Rb$^+$ ions normal to the threefold axis $C_3$.
Within the layers, the ions form regular triangular lattices.
Figure~\ref{fig2_crystal} shows the magnetic Fe$^{3+}$ ions (3$d^5$, $S=5/2$) and the non-magnetic Rb$^+$ ions, the (MoO$_4$)$^{2-}$ complexes are omitted for clarity.~\cite{Klevtsova_1970}
The lattice parameters of \rfm\ are $a=5.69$~\AA\ and $c=7.48$~\AA.
At room temperature, the crystal structure  belongs to the space group $P\bar{3}m1$ and below 190~K the symmetry is lowered to the space group $P\bar{3}c1$.~\cite{Klimin_2003}
The in-plane exchange integral $J=0.086(2)$~meV is antiferromagnetic and is much stronger than the inter-planar integrals $J'$, $J_a$, $J_b$.~\cite{White_2013}
The inter-plane exchange integral is approximately one hundred times smaller than the in-plane exchange integral, $J'/J=0.008(1)$ according to neutron diffraction experiments and $J'/J=0.01$ according to ESR measurements.~\cite{White_2013,Svistov_2003,Svistov_2006_Err}
The ratio of the diagonal inter-plane exchange integrals was also evaluated from first principle calculations~\cite{KunCao_2014} as $|J_{a,b}| / J \approx 0.002$.

\rfm\ orders magnetically below $T_{\rm N}\approx 3.8$~K~\cite{Inami_1996}.
The magnetic structures are found for the in-plane field orientation using a variety of experimental
techniques.~\cite{Inami_1996,Svistov_2003,Kenzelmann_2004,Svistov_2006,Kenzelmann_2007,Smirnov_2007,White_2013,Mitamura_2014,Zelenskiy_2021} The description of the magnetic states of the H-T phase diagram obtained in these experiments are discussed in detail in Ref.~[\onlinecite{LAST}].

\begin{figure}[tb]
\includegraphics[width=0.95\columnwidth]{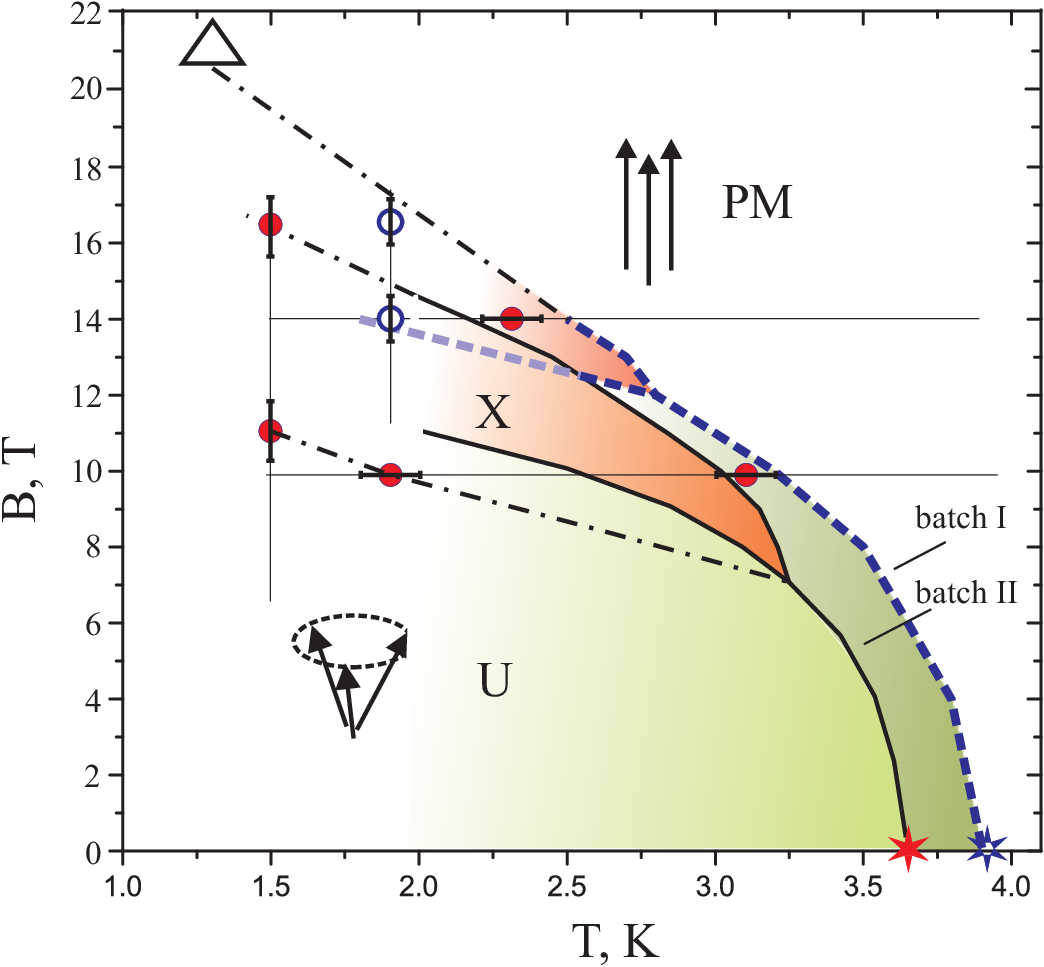}
\caption{Magnetic phase diagram of \rfm\ for $H\parallel C_3$. Magnetic phases are labeled U-phase (green), unknown high field X phase (red), and polarized PM phase.
The dashed lines in the figure mark the phases boundaries from Refs.~[\onlinecite{Smirnov_2007, Mitamura_2016}] (see text). The solid line show boundaries of phases found by authors of Ref.~[\onlinecite{Zelenskiy_2021}] from acoustic experiments.
The thin solid lines show the temperature and field scans made in the
NMR study. The dash-dotted lines are guide for the eyes.
The open blue circles and solid red circles mark the transitions between U and X or X and PM phases obtained from the NMR experiments on a sample from batches I and II respectively.
The red and blue stars show the $T_{\rm N}$ measured in samples of batches I and II from the temperature dependence of the magnetization at $\mu_{0}H=0.1$~T.
The open triangle shows the $H_{sat}$ measured from $M(H)$ in pulse magnetic field on the sample from batch I.~\cite{Smirnov_2007}}
\label{fig3_PhD}
\end{figure}

The H-T phase diagram for H aligned perpendicular to triangular planes is shown in Fig.~\ref{fig3_PhD}. The dashed lines in the figure mark the phase boundaries from Refs.~[\onlinecite{Smirnov_2007, Mitamura_2016}] obtained from anomalies on temperature and field  dependencies of magnetic moment, heat capacity, electric polarization and dielectric constant. The solid line show boundaries of phases found by authors of Ref.~[\onlinecite{Zelenskiy_2021}] from acoustic experiments. The phase diagrams obtained in two different samples are generally similar, with the Neel temperature differing slightly by 0.25~K. A scatter in $T_N$ within 0.3~K was observed previously in samples from different batches. The samples with smaller $T_N$ demonstrate reduced values of the saturation field as well as the transition from U to X phases occurs at lower fields. In this work we studied the H-T evolution of NMR spectra and spin-lattice relaxation time $T_1$ along dotted lines shown in the diagram. We studied two single crystals with slightly different Neel temperatures. The points corresponding to anomalies in $T_1(T,H)$ at phase transitions and $T_N$ are depicted with open (blue) and closed (red) symbols for samples with higher and lower Neel temperatures respectively. The open square symbol for sample with higher $T_N$ denotes the saturation field measured in pulse field.~\cite{Smirnov_2007, Mitamura_2016} Dash-dotted lines are guides to the eye. The H-T range of low field U phase is colored green. This phase is characterized by presence of spontaneous electrical polarization directed along $C_3$.
In zero field, a 120$^{\circ}$ magnetic structure is established within every triangular plane.
The propagation vector of the incommensurate structure is  (1/3,~1/3,~0.458).~\cite{Kenzelmann_2007}

\section{Sample preparation and experimental details}

We used two different batches of \rfm\ single crystals, as in our previous work~\cite{LAST}, which reports the results of $^{87}$Rb NMR for $H\perp C_3$.
The batches are labeled as I and II following Ref.~[\onlinecite{LAST}].
Powder x-ray diffraction study of the structures did not show a significant difference between the batches, however, the magnetic ordering temperature $T_{\rm N}$ and the values of $H_{\rm sat}$ for the second batch were about 7\% lower than for the samples from the first batch.
A typical size of the crystal was $2 \times 2 \times 0.5$~mm$^3$ with the smallest dimension corresponding to the $C_3$ direction of the crystal.

NMR measurements were taken in a superconducting Cryomagnetics 17.5~T magnet at the National High Magnetic Field Laboratory, Florida, USA.
$^{87}$Rb nuclei (nuclear spin $I=3/2$, gyromagnetic ratio $\gamma/2\pi=13.9318$~MHz/T) were probed using a pulsed NMR technique.
The spectra were obtained by summing fast Fourier transforms (FFT) spin-echo signals as the field was swept through the resonance line.
Utilizing FFT techniques, NMR spin echoes were obtained using $\tau_p - \tau_D - 2\tau_p$ pulse sequences, where the pulse lengths $\tau_p$ were 1~$\mu$s and the times between pulses $\tau_D$ were 15~$\mu$s.
$T_1$ was extracted using a multiexponential expression which is utilized in spin-lattice relaxation when NMR lines are split by quadrupole interaction~\cite{Suter_1998}.
The measurements were carried out in the temperature range $1.43 \leq T \leq 25$~K, temperature stability was better than 0.05~K.

\section{Experimental results}

\begin{figure}[tb]
\includegraphics[width=0.8\columnwidth]{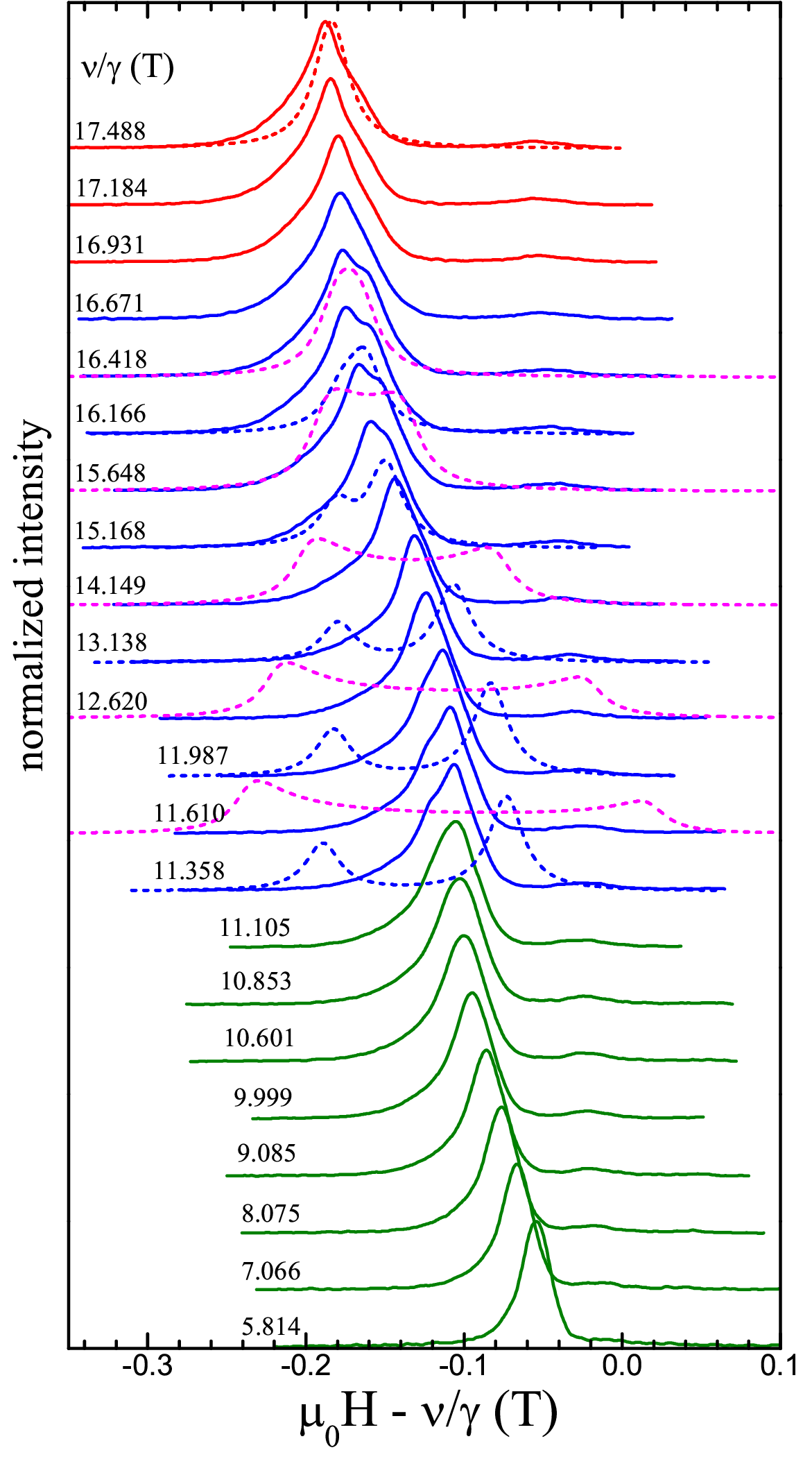}
\caption{$^{87}$Rb NMR spectra of the transition  $m_I = -1/2 \leftrightarrow +1/2$ measured at $T=1.45$~K in the range from 5.8 to 17.5~T with the field applied parallel to $C_3$~axis of the crystal on the sample from batch II.
Lines measured at different frequencies are offset for clarity.
Red lines are the spectra measured in a field-polarized PM phase.
Blue lines are the spectra obtained in fields below $H_{\rm sat}$ where the X phase is expected.
The spectra in low fields shown in green are obtained in the U phase.
The boundary fields between phases was found from lambda anomaly on $T_1(H)$ dependency presented in Figure ~\ref{fig5_NMR_T1_H}.
The dashed lines are the model spectra computed for PM(red), V(blue) and fan(magenta) structures using iron magnetic moment $5~\mu_B/Fe^{3+}$, individual linewidth $\delta=13$mT (see text).
}
\label{fig4_NMR_all}
\end{figure}

\begin{figure}[tb]
\includegraphics[width=0.95\columnwidth]{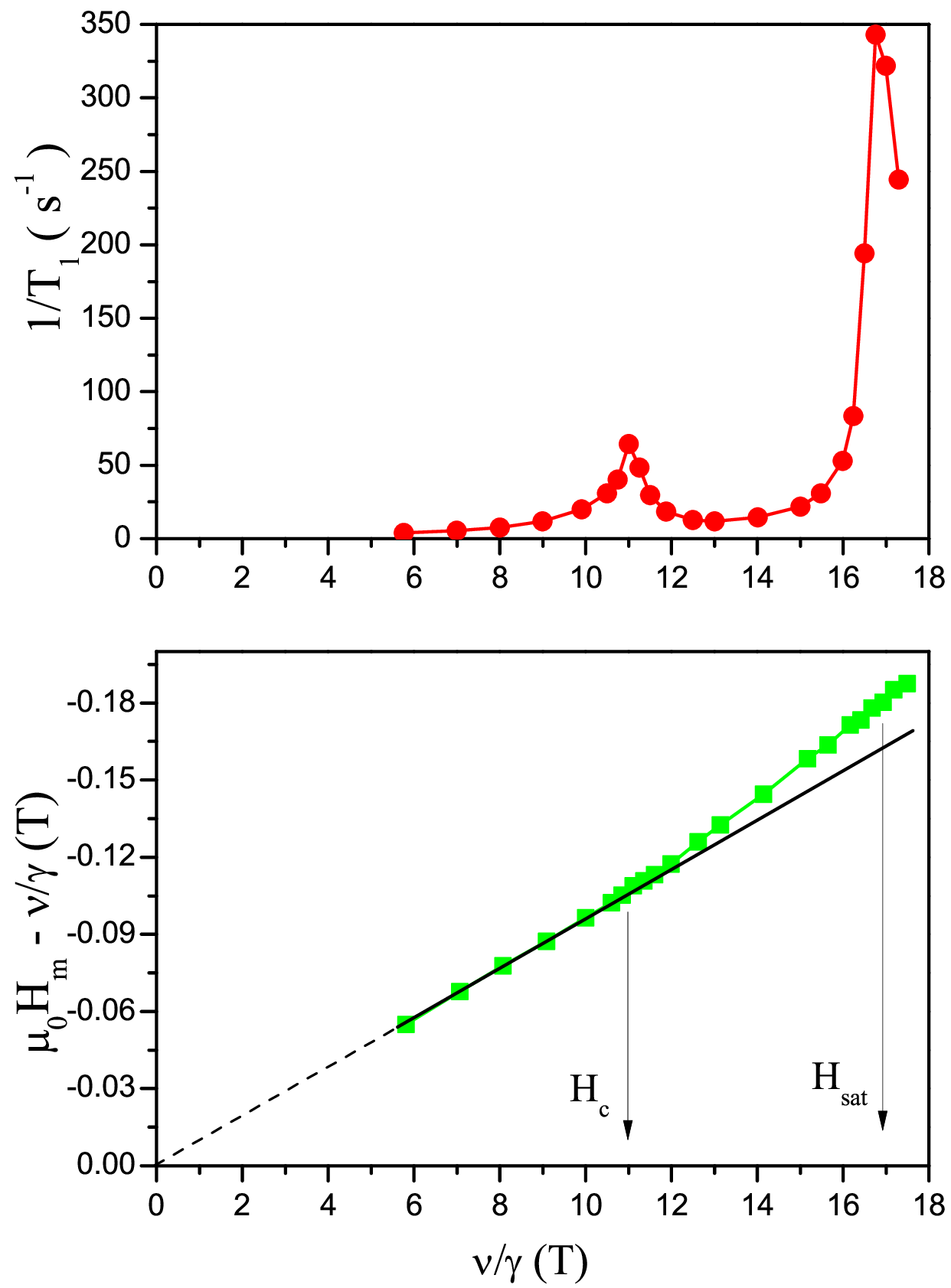}
\caption{Field dependencies of (a) the spin-lattice relaxation rate, $1/T_1$, and (b) the effective field $\mu_0 H_m - \nu/\gamma$ of $^{87}$Rb NMR measured at $T=1.45$~K. The resonance fields $H_m$ used in (b) were derived from the center of gravity of each spectra shown in Figure~\ref{fig4_NMR_all}. }
\label{fig5_NMR_T1_H}
\end{figure}

Figure~\ref{fig4_NMR_all} shows the central line of the $^{87}$Rb NMR spectra corresponding to the central transition $m_I = -1/2 \leftrightarrow +1/2$. The two quadrupole split satellites corresponding to the transitions $m_I = \pm 3/2\leftrightarrow \pm 1/2$ are not shown. The spectra are measured at $T = 1.45$~K in the field range $5.8 < \mu_0 H <17$~T with the field applied parallel to $C_3$~axis of the crystal. In the polarized PM phase, all the Rb$^+$ ions are in equivalent positions and the spectrum consists of one line. In the magnetically ordered phase, the positions of the Rb$^+$ ions are not equivalent and, as a result, the local magnetic fields from the magnetic neighbors is expected to be different and each line of the quadrupole split spectra at $H <H_{\rm sat}$ would exhibit a complex structure.
The effective fields from the ordered components of the magnetic moments are smaller than the quadrupolar splitting which allows for the observation of the fine structure on every satellite without overlapping.~\cite{Svistov_2005, LAST} This is the reason why we only show the central line of the spectra corresponding to the transition $m_I = -1/2\leftrightarrow +1/2$.
The obtained spectra are shown in Fig.~\ref{fig4_NMR_all}  against $\mu_0 H-\nu/\gamma$, where $\nu$ is frequency. Colors are used to mark the spectra obtained in different magnetic phases, PM (red), X (blue), U (green). The boundary fields between phases were found from lambda anomaly on $T_1(H)$ dependency presented in the panel (a) Figure~\ref{fig5_NMR_T1_H}.

According Figure~\ref{fig4_NMR_all} we can conclude that a single line $^{87}$Rb NMR spectrum in full field range is observed. Line broadening  and fine structure on the lineshapes observed at higher fields can be ascribed to distribution of demagnetization fields in the  sample of non-elliptical shape.
Neglecting the variation of the spin-spin relaxation time, $T_2$, within the NMR spectrum, the measured intensity of the echo signal is proportional to the number of rubidium nuclei in the effective field from the magnetic neighbors with projection on $H$ equal to $\mu_0 H-\nu/\gamma$. The panel (b) of Figure~\ref{fig5_NMR_T1_H} shows the field dependence of effective field $\mu_0 H-\nu/\gamma$ on $^{87}$Rb. The resonance fields $H_m$ used in this graph were derived from the center of gravity for each spectra shown in Figure~\ref{fig4_NMR_all}. The field dependency of effective field on $^{87}$Rb is linear up to the transition field $H_c$ from U to X phase where the slope increases approximately $25~\%$.

\begin{figure}[tb]
\includegraphics[width=0.8\columnwidth]{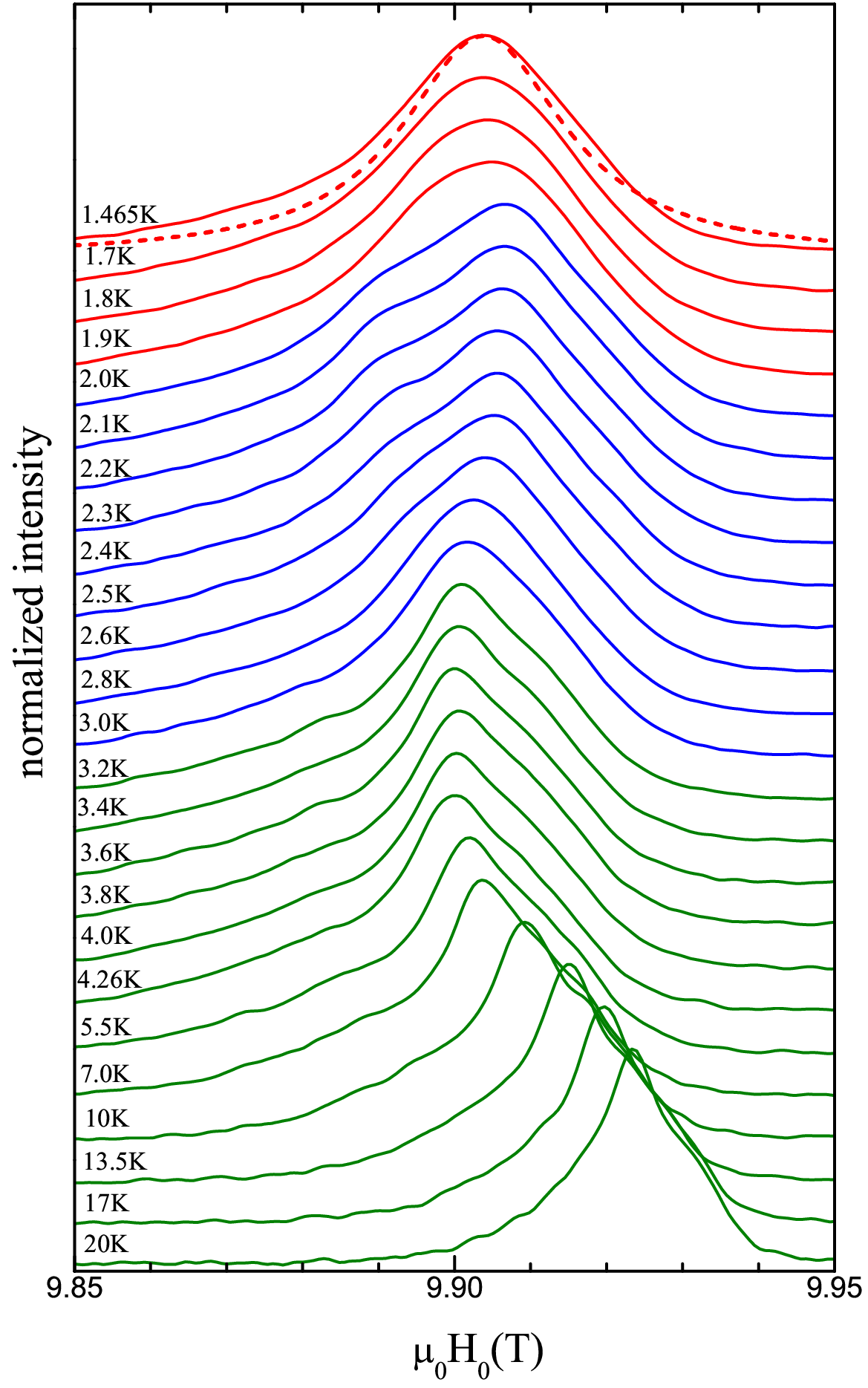}
\caption{$^{87}$Rb NMR spectra of the transition  $m_I = -1/2 \leftrightarrow +1/2$ measured at $\nu=139.3$~MHz in the temperature range from 1.5 to 20~K with the field applied parallel to $C_3$~axis of the crystal on the sample from batch II.
Lines measured at different temperatures are offset for clarity.
Red, blue, green colours correspond to spectra measured in the U, X and PM phases, respectively.
The boundary fields between phases was found from lambda anomaly on $T_1(H)$ dependency presented in Figure ~\ref{fig7_NMR_T1_T}.
The dashed line is the model spectrum computed for U phase using iron magnetic moment $2.4~\mu_B/Fe^{3+}$, individual linewidth $\delta=13$mT (see text).
}
\label{fig6_NMR_all_T}
\end{figure}

\begin{figure}[tb]
\includegraphics[width=0.9\columnwidth]{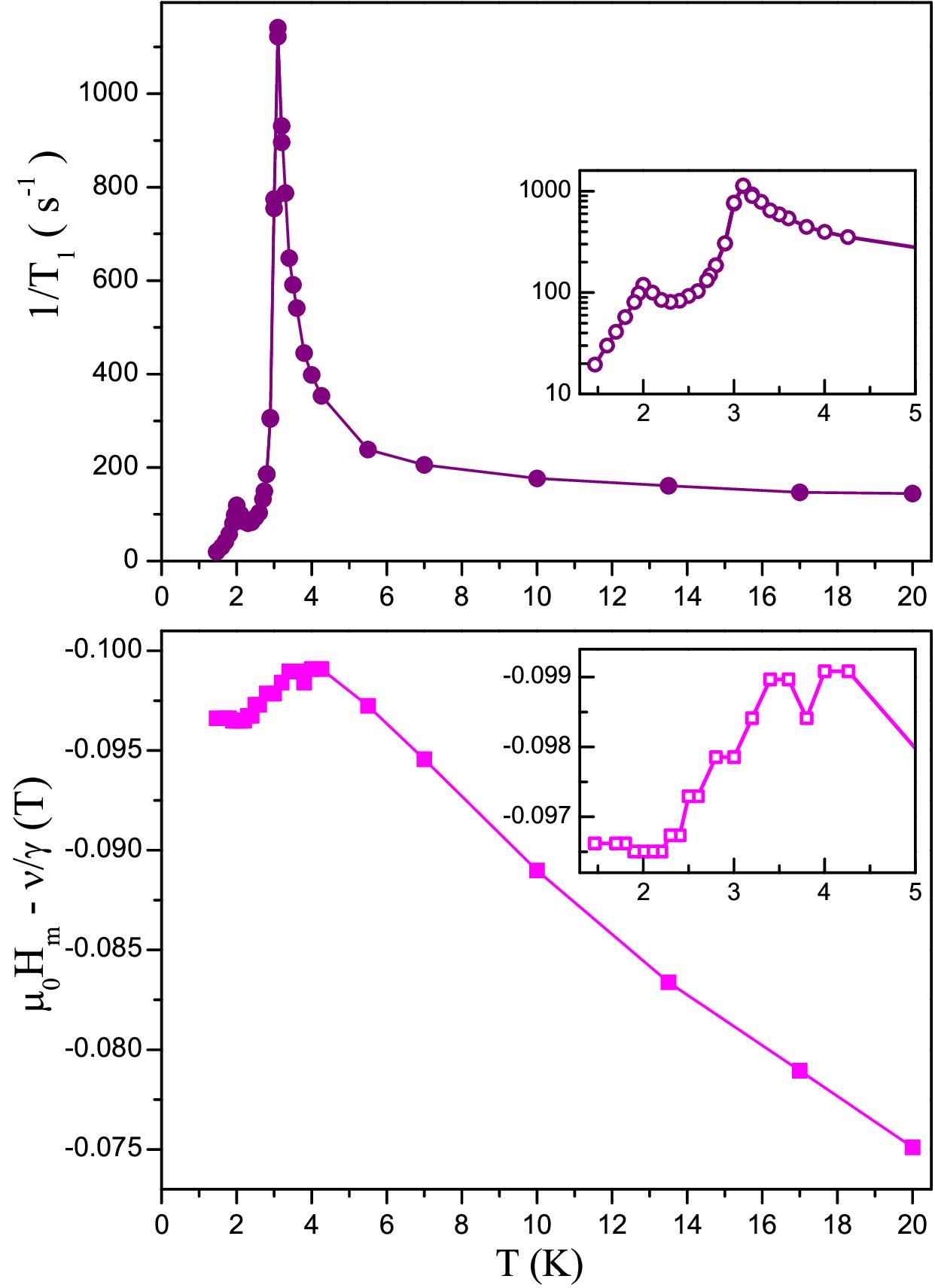}
\caption{Temperature dependencies of (a) the spin-lattice relaxation rate, $1/T_1$, and (b) the effective field $\mu_0 H_m - \nu/\gamma$ of $^{87}$Rb NMR measured at $\nu=139.3$~MHz. The resonance field $H_m$ used in b) was found as the center of gravity for measured spectra shown in Figure~\ref{fig6_NMR_all_T}.
Insets repeat the main panels in enlarged scales.
}
\label{fig7_NMR_T1_T}
\end{figure}

Figure~\ref{fig6_NMR_all_T} shows the NMR spectra measured at 139.3 MHz, $H\parallel C_3$ at different temperatures. Colors are used to mark the spectra obtained in different magnetic phases, PM (red), X (blue),  U (green).
In the whole temperature range the NMR spectra could be considered also as single line spectra. The boundary fields between phases were identified from lambda anomaly on $T_1(T)$ dependency presented in   Figure~\ref{fig7_NMR_T1_T}(a). The temperature dependence of effective field $\mu_0 H_m - \nu/\gamma$ on $^{87}$Rb is shown in Figure~\ref{fig7_NMR_T1_T}(b). Transition temperatures identified from lambda anomaly are shown with arrows.

Concluding, we can say, first, that the high field X magnetic phase for  $H\parallel C_3$ of \rfm\ below saturation field was observed on samples of two different batches. Secondly, the $^{87}$Rb NMR spectra observed for this direction have a single lineshape in all studied H-T range. This result implies that within U, X and PM structures the projections of effective field on rubidium nuclei on H are the same.

\section{Discussion}

The magnetic resonance condition of $^{87}$Rb nuclei in our experiments is defined by an interaction with a strong external magnetic field $\nu=\gamma H$, with small corrections due to the quadrupole interaction, dipolar interaction with the magnetic environment and transferred hyperfine interactions with the nearest magnetic neighbors.
The corrections to the resonance field due to quadrupolar interaction are the same for all $^{87}$Rb nuclei and, as a result, they lead to a field shift of the NMR spectra.
In a high-field range, this shift for the central NMR line $m_I = -1/2\leftrightarrow +1/2$ which was studied in our experiments is negligibly small.~\cite{Svistov_2005}
The largest contribution to the effective field is dipolar, which can be directly computed for the different magnetic structures.

The dipolar fields were calculated for a $^{87}$Rb nucleus located in the middle of a cylindrical sample with the base radius of 100$a$ and the height of 10$c$.
Such shape of the model sample takes the demagnetizing field into account.

The model spectra for the frequency 243.633~MHz ($\nu/\gamma=17.5$~T) is shown in the Figure~\ref{fig4_NMR_all} with the red dashed line. It is obtained assuming the magnetic moment on iron ions equals $5~\mu_B/Fe^{3+}$. The model line is closely fits the experimental one.

The modeling of NMR spectra in magneto ordered phases follows the same procedure as described in previous works (Refs.~[\onlinecite{LAST,Sakhratov_2019}]).
In the of case of magnetic structures with the wave vectors $(1/3, 1/3, k_z)$, the dipolar fields from the Fe$^{3+}$ ions acting on the $^{87}$Rb nuclei are caused only by the components of the magnetic moments parallel to the applied field, as the dipolar fields from the components perpendicular to the field cancel each other. The projections of the magnetic moments of three sublattices along the static field H for U structure are the same. Thus for U phase is expected to be a single line shaped NMR spectra.
The red dashed line shown in the Figure~\ref{fig6_NMR_all_T} at $T=1.5$~K is the NMR spectra computed in the model U phase with the magnetic moment of Fe$^{3+}$ ions projected on the field direction equals to $2.4~\mu_B/Fe^{3+}$. The value of magnetic moment can be found from the M(H) measurements at 1.6~K given in (Fig.~3), Ref.~[\onlinecite{Svistov_2003}].

Let us consider a possibility of a V structure in the X phase as suggested in Refs.~[\onlinecite{Mitamura_2016, Zelenskiy_2021}]. The computed spectra for planar commensurate V structure with wave vector (1/3, 1/3, 1/3) are shown in Fig.4 as blue dashed lines. For this case, we expect two lines, with the intensity of high field line twice as large as the intensity of the low field line.\cite{LAST} The model spectra in Fig. 4 were computed assuming the value of magnetic moment on Fe$^{3+}$ ion equals $5\mu_B$. The splitting of the two-peak model spectra decreases with field and disappears at saturated field. Our experimental spectra do not show any appreciable change during the transition from U to X phase. Thus we can exclude the V structure from consideration as an option for the X phase.

In Ref.~[\onlinecite{LAST}], we have identified the magnetic structure for $H~\perp~C_3$ below saturation as an incommensurate fan-phase rather than a commensurate V-phase.
The presence of this phase was determined with NMR and neutron diffractions experiments. An
analysis of the NMR spectra shows that the high-field fan phase of \rfm\ can be successfully described
by a periodic commensurate oscillation of the magnetic moments around the field direction in each Fe$^{3+}$ layer
combined with an incommensurate modulation of the magnetic structure perpendicular to the layers. We computed the $^{87}$Rb NMR spectra for the model fan structure with spin orientations on the iron ions given in Ref.~[\onlinecite{LAST}]. The calculated spectra for the case $H~\parallel~C_3$ demonstrate broad line with two maxima at the boundaries usual for the incommensurate structures. These spectra are calculated in the V model and are shown in Fig.~4 as dashed magenta lines. Here, we assume that the incommensurate vector of the structure  is (1/3, 1/3, $k_{ic}$), with $k_{ic} \approx$~0.5, as it was found experimentally in the U phase.~\cite{Kenzelmann_2004} Thus, modeling the NMR spectra in terms of either V or fan structure is inconsistent with the observed single line experimental data.

Finally we summarize the main features of X-phase known from literature and obtained in the present work.
1. The boundaries of X phase are definitely detected  as anomalies on  field dependencies of dielectric constant~\cite{Mitamura_2016}, sound velocities~\cite{Zelenskiy_2021} and the spin-lattice relaxation time $T_1$ of $^{87}$Rb NMR.
2. The electric polarization, specific for U phase, is not observed for X phase.~\cite{Mitamura_2016}
3. The projections of the effective fields from magnetic neighbors along the direction of $\vect{H}_0$ at all rubidium
ions are the same. This result agrees with U and PM structures and is inconsistent with V and fan structures.
4. Knight shift of $^{87}$Rb, proportional to magnetization of \rfm\, demonstrates inflection in the field dependency during the transition from U-phase to X-phase.
This shift can be explained by $\approx 25~\%$ increase of magnetic susceptibility at this field.
Surprisingly, differential susceptibility $dM/dH$ does not demonstrate any perceptible change during the transition from U to V phase.~\cite{Smirnov_2007, Mitamura_2016} Possibly this difference is due to non equilibrium magnetic state during pulsed fields measurements.

As a result, we suggest two possible magnetic structures for X phase which are in agreement with the observed NMR spectra.
1.~Umbrella-like magnetic structure in every individual triangular plane with inter plane disorder, and~2. Spin structure with tensor order parameter within every triangular plane.
For both structures, the projections of the moments along the field direction are the same for all Fe$^{3+}$ ions and as a result, a single line in $^{87}$Rb NMR spectra is expected which would be in agreement with the experiment.

\acknowledgments
We thank A.I.~Smirnov, S.S.~Sosin, A.V. Syromyatnikov, and M.E.~Zhitomirskiy for stimulating discussions.
The modeling of NMR spectra was supported by the Russian Science Foundation Grant No. 22-12-00259.
Work at the National High Magnetic Field Laboratory is supported by the User Collaborative Grants Program (UCGP) under NSF Cooperative Agreement No.~DMR-1157490 and DMR-1644779, and the State of Florida.


\begin{thebibliography}{}
\bibitem{Inami_1996} T.~Inami, Y.~Ajiro, and T.~Goto, J. Phys. Soc. Jpn. {\bf{65}}, 2374 (1996).
\bibitem{Svistov_2003} L.~E.~Svistov, A.~I.~Smirnov, L.~A.~Prozorova, O.~A.~Petrenko, L.~N.~Demianets, and A.~Ya.~Shapiro, Phys. Rev. B {\bf{67}}, 094434 (2003).
\bibitem{Kenzelmann_2004}  G.~A.~Jorge, C.~Capan, F.~Ronning, M.~Jaime, M.~Kenzelmann, G.~Gasparovic, C.~Broholm, A.~Ya.~Shapiro, L.~N.~Demianets, Physica B {\bf{354}}, 297 (2004).
\bibitem{Svistov_2006} L.~E.~Svistov, A.~I.~Smirnov, L.~A.~Prozorova, O.~A.~Petrenko, A.~Micheler, N.~B\"uttgen, A.~Y.~Shapiro, and L.~N.~Demianets, Phys. Rev. B {\bf{74}}, 024412 (2006).
\bibitem{Kenzelmann_2007} M.~Kenzelmann, G.~Lawes, A.~B.~Harris, G.~Gasparovic, C.~Broholm, A.~P.~Ramirez, G.~A.~Jorge, M.~Jaime, S.~Park, Q.~Huang, A.~Ya.~Shapiro, and L.~A.~Demianets, Phys. Rev. Lett. {\bf{98}}, 267205 (2007).
\bibitem{Smirnov_2007} A.~I.~Smirnov, H.~Yashiro, S.~Kimura, M.~Hagiwara, Y.~Narumi, K.~Kindo, A.~Kikkawa, K.~Katsumata, A.~Ya.~Shapiro, and L.~N.~Demianets, Phys. Rev. B {\bf{75}}, 134412 (2007).
\bibitem{White_2013} J.~ S.~White, Ch.~Niedermayer, G.~Gasparovic, C.~Broholm, J.~M.~S.~Park, A.~Ya.~Shapiro, L.~A.~Demianets, and M.~Kenzelmann, Phys. Rev. B {\bf{88}}, 060409(R) (2013).
\bibitem{Mitamura_2014} H.~Mitamura, R.~ Watanuki, K.~Kaneko, N.~Onozaki, Y.~Amou, S.~Kittaka, R.~Kobayashi, Y.~Shimura, I.~Yamamoto, K.~Suzuki, S.~Chi, and T.~Sakakibara, Phys. Rev. Lett. {\bf{113}}, 147202 (2014).
\bibitem{Zelenskiy_2021} A.~Zelenskiy, J.~A.~Quilliam, A.~Ya.~Shapiro, and G.~Quirion, Phys. Rev. B {\bf 103}, 224422 (2021).
\bibitem{Korshunov_1986} S.~E.~Korshunov, J. Phys. C: Solid State Phys. {\bf{19}}, 5927 (1986).
\bibitem{Lee_1986} D.~H.~Lee, J.~D.~Joannopoulos, J.~W.~Negele, D.~P.~Landau, Phys. Rev. B {\bf{33}}, 450 (1986).
\bibitem{Chubukov_1991} A.~V.~Chubukov, D.~I.~Golosov, J. Phys.: Condens. Matter {\bf{3}}, 69 (1991).
\bibitem{Gekht_1997} R.~S.~Gekht and I.~N.~Bondarenko, J. Exp. Theor. Phys. {\bf 84}, 345 (1997).
\bibitem{LAST} Yu.~A.~Sakhratov, O.~Prokhnenko , A.~Ya.~Shapiro, H.~D.~Zhou, L.~E.~Svistov , A.~P.~Reyes, and O.~A.~Petrenko, Phys. Rev. B {\bf{105}}, 014431 (2022).
\bibitem{Mitamura_2016} H.~Mitamura, R.~Watanuki, N.~Onozaki, Y.~Amoub, Y.~Kono, S.~Kittaka, Y.~Shimura, I.~Yamamoto, K.~Suzuki, and T.~Sakakibara, Journal of Magnetism and Magnetic Materials {\bf 400}, 70 (2016).
\bibitem{Klevtsova_1970} R.~F.~Klevtsova and P.~V.~Klevtsov, Kristallografiya {\bf 15}, 953 (1970).
\bibitem{Klimin_2003} S.~A.~Klimin, M.~N.~Popova, B.~N.~Mavrin, P.~H.~M.~van~Loosdrecht, L.~E.~Svistov, A.~I.~Smirnov, L.~A.~Prozorova, H.-A.~Krug~von~Nidda, Z.~Seidov, A.~Loidl, A.~Ya.~Shapiro, and L.~N.~Demianets, Phys. Rev. B {\bf 68}, 174408 (2003).
\bibitem{Svistov_2006_Err}  L.~E.~Svistov, A.~I.~Smirnov, L.~A.~Prozorova, O.~A.~Petrenko, L.~N.~Demianets, and A.~Y.~Shapiro, Phys. Rev. B  {\bf 74}, 139901(E) (2006).
\bibitem{KunCao_2014} Kun Cao, R.~D.~Johnson, Feliciano Giustino, P.~G.~Radaelli, G-C.~Guo, and Lixin He, Phys. Rev. B {\bf 90}, 024402 (2014).
\bibitem{Suter_1998} A.~Suter, M.~Mali, J.~Roos, and D.~Brinkmann, J. Phys.: Condens. Matter {\bf{10}}, 5977 (1998).
\bibitem{Svistov_2005} L.~E.~Svistov, L.~A.~Prozorova,  N.~B\"uttgen, A.~Ya.~Shapiro, and L.~N.~Demianets, JETP Lett. {\bf 81}, 102 (2005).
\bibitem{Sakhratov_2019} Yu.~A.~Sakhratov, M.~Prinz-Zwick, D.~Wilson, N.~Buuttgen, A.~Ya.~Shapiro, L.~E.~Svistov, and A.~P.~Reyes, Phys. Rev. B {\bf 99}, 024419 (2019).



\end{thebibliography}
\end{document}